\documentclass[english]{article}
\usepackage[T1]{fontenc}
\usepackage[latin9]{inputenc}
\usepackage[active]{srcltx}
\usepackage{color}
\usepackage{float}
\usepackage{amsmath}
\usepackage{amssymb}
\usepackage{graphicx}

\makeatletter
\usepackage{amsfonts,amssymb}
\usepackage{latexsym}
\usepackage{epsfig}
\usepackage{cite}
\usepackage{graphicx}
\usepackage[colorlinks,linkcolor=blue]{hyperref}
\newcommand{\be}{\begin{equation}}
\newcommand{\ee}{\end{equation}}

\usepackage{tikz}

\makeatother

\usepackage{babel}
\begin{document}
{}~ \hfill\vbox{\hbox{ }}\break
\vskip 3.0cm
\centerline{\Large \bf Probing black hole entropy via entanglement} 
\vspace*{10.0ex}

\vspace*{10.0ex}
\centerline{\large Shuxuan Ying}
\vspace*{7.0ex}
\vspace*{4.0ex}

\centerline{\large \it Department of Physics}
\centerline{\large \it Chongqing University}
\centerline{\large \it Chongqing, 401331, China} \vspace*{1.0ex}

\vspace*{4.0ex}
\centerline{\large  ysxuan@cqu.edu.cn}

\vspace*{4.0ex}
\centerline{\bf Abstract} \bigskip \smallskip
In this paper, we develop a method to extract the Bekenstein-Hawking entropy of  $D$-dimensional black holes using the entanglement entropy of a lower-dimensional conformal field theory (CFT). This approach relies on two key observations. On the gravitational side, the near-horizon geometry of extremal black holes is AdS$_{2}$, and the Bekenstein-Hawking entropy is entirely determined by this two-dimensional geometry. Moreover, the higher-dimensional spherical part of the black hole metric is absorbed into the $D$-dimensional Newton's constant $G_{N}^{\left(D\right)}$, which can be effectively reduced to a two-dimensional Newton's constant  $G_{N}^{\left(2\right)}$. On the field theory side, the entanglement entropy of two disconnected one-dimensional conformal quantum mechanics (CQM$_{1}$) can be calculated. According to the Ryu-Takayanagi (RT) prescription, this entanglement entropy computes the area of the minimal surface in the AdS$_{2}$ geometry. Since the near-horizon region of the black hole and the emergent spacetime derived from the entanglement entropy share the same Penrose diagram---with both the black hole event horizon and the RT surface corresponding to specific points on this diagram---the Bekenstein-Hawking entropy can be probed via entanglement entropy when these points coincide. This result explicitly demonstrates that the entanglement across the event horizon is the fundamental origin of the Bekenstein-Hawking entropy.

\vfill 
\eject
\baselineskip=16pt
\vspace*{10.0ex}
\tableofcontents

\section{Introduction}

Understanding the deep connection between the Bekenstein-Hawking entropy
of black holes and entanglement entropy is crucial. It provides insights
into the microscopic origin of black hole entropy and offers a pathway
toward resolving the black hole information paradox. In general, quantum
corrections to the Bekenstein-Hawking entropy are interpreted as entanglement
entropy contributions \cite{Susskind:1994sm,Fiola:1994ir}. However,
if gravity itself is entirely induced, the Bekenstein-Hawking entropy
may be identified directly with entanglement entropy \cite{Susskind:1994sm,Jacobson:1994iw}.
In this paper, motivated by the work \cite{Azeyanagi:2007bj}, we
aim to probe higher-dimensional Bekenstein-Hawking entropy through
the entanglement entropy of a lower-dimensional conformal field theory
(CFT), based on the equivalence:

\begin{equation}
S_{EE}=\frac{\mathrm{Area}\left(\gamma_{A}\right)}{4G_{N}^{\left(d+2\right)}}=\frac{\mathrm{Area}\left(\Sigma\right)}{4G_{N}^{\left(d+2\right)}}=S_{BH}.
\end{equation}

\noindent which holds when the area of the $d$-dimensional minimal
surface $\gamma_{A}$, as given by the Ryu--Takayanagi (RT) formula,
coincides with the area of the black hole event horizon $\Sigma$.
Although the possibility of such an equivalence was also explored
in ref.\,\cite{Emparan:2006ni}, through the study of the Emparan--Horowitz--Myers
black hole \cite{Emparan:1999wa}, our method applies more generally
to arbitrary extremal black holes.

This equivalence is most transparent in three dimensions, where both
$\mathrm{Area}\left(\gamma_{A}\right)$ and $\mathrm{Area}\left(\Sigma\right)$
reduce to geodesic lengths and can be directly identified. A well-known
example is the BTZ black hole \cite{Azeyanagi:2007bj}. If we consider
its boundary as a circle parameterized by $\phi\sim\phi+2\pi$, and
define entanglement entropy on this boundary between two regions---region
$A$ with angular size $\triangle\phi=2\pi L$, and region $B$ with
$\triangle\phi=2\pi\left(1-L\right)$---then the entanglement entropy
takes the form

\begin{equation}
S_{A}\left(L\right)=\frac{c}{3}\log\left[\frac{\beta}{\pi a}\sinh\left(\frac{\pi L}{\beta}\right)\right],
\end{equation}
where $\beta$ is the inverse temperature of the black hole, and $c$
is the central charge of the CFT on the boundary. In the high-temperature
limit, as $L\rightarrow1$, we let $L=1-\epsilon$ with $\epsilon\rightarrow0$.
Then, the entanglement entropy becomes $S_{A}\left(L=1-\epsilon\right)=S_{BH}+S_{A}\left(L=\epsilon\right)$
indicating that the black hole entropy $S_{BH}$ can be extracted
from the entanglement entropy. Holographically, the entanglement entropy
$S_{A}\left(L\right)$ corresponds to the geodesic length of $\gamma_{A}$
via

\begin{equation}
S_{A}\left(L\right)=\frac{\mathrm{Length}\left(\gamma_{A}\right)}{4G_{N}^{\left(3\right)}},\qquad\mathrm{with}\qquad c=\frac{3R_{AdS}}{2G_{N}^{\left(3\right)}}.
\end{equation}

\begin{figure}[H]
\begin{centering}
\includegraphics[scale=0.4]{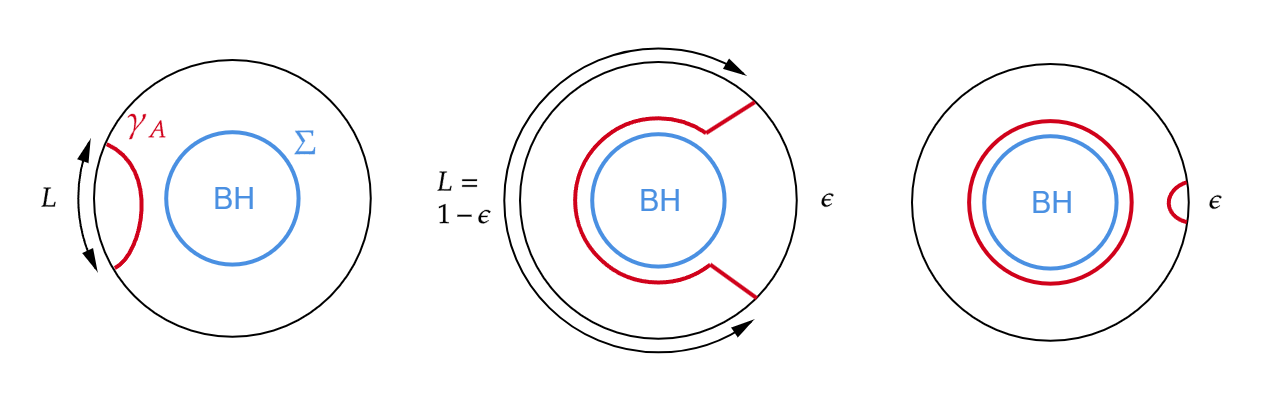}
\par\end{centering}
\caption{\label{fig:BTZ_EE}The geodesic length of $\gamma_{A}$ on a time
slice of AdS$_{3}$ is related to the entanglement entropy of the
dual CFT$_{2}$ for an entangling region of length $L$. As $L$ increases,
the geodesic $\gamma_{A}$ stretches further and eventually wraps
around the event horizon of the black hole. In the limit where $L$
becomes sufficiently large, the geodesic $\gamma_{A}$ fully encircles
the horizon, and its length coincides with the horizon length. Consequently,
the black hole entropy can be extracted from the entanglement entropy
in this manner.}
\end{figure}

\noindent As $L$ increases, the geodesic $\gamma_{A}$ winds around
the BTZ black hole horizon, and its length approaches the horizon
circumference, i.e., $\mathrm{Length}\left(\gamma_{A}\right)=\mathrm{Area}\left(\Sigma\right)$,
as illustrated in figure (\ref{fig:BTZ_EE}). This demonstrates that
the black hole entropy can indeed be obtained from entanglement entropy.

\begin{figure}[H]
\begin{centering}
\includegraphics[scale=0.35]{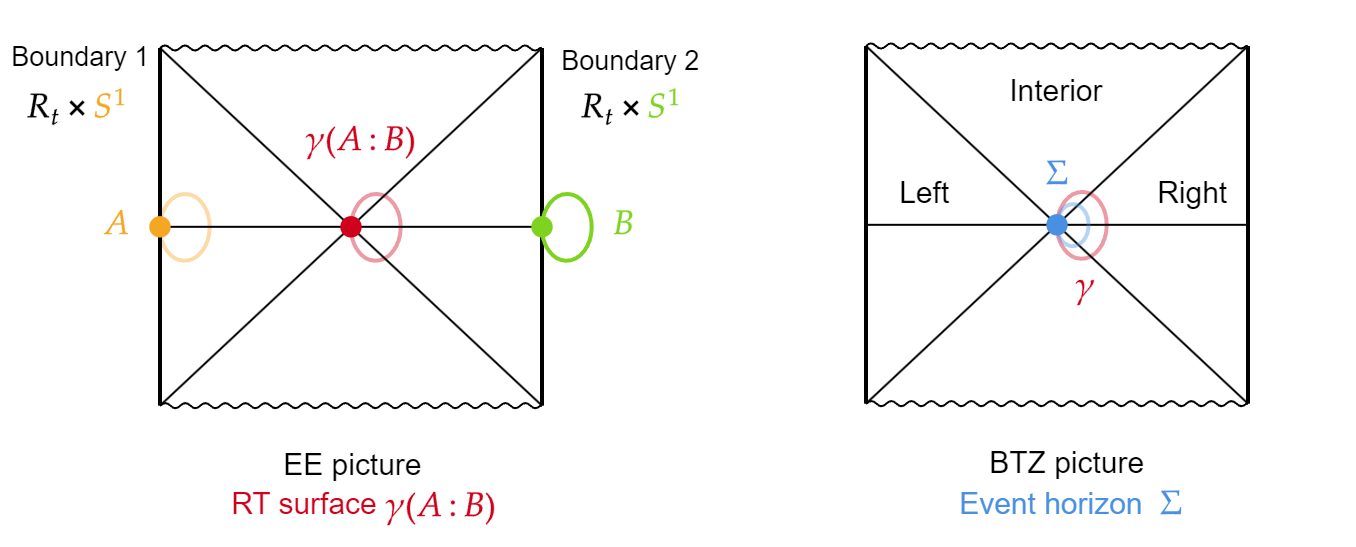}
\par\end{centering}
\caption{\label{fig:BTZ_EE2}The left panel illustrates the emergence of spacetime
from the entanglement entropy of a thermofield double (TFD) state.
Two CFTs reside on the two asymptotic boundaries, and at a fixed time
slice, each point on the boundary corresponds to an $S^{1}$ circle.
The entanglement entropy between two disconnected boundary circles
yields the length of the geodesic ${\color{red}\gamma\left(A:B\right)}$.
This geodesic ${\color{red}\gamma\left(A:B\right)}$ is precisely
the event horizon ${\color{blue}\Sigma}$ of the BTZ black hole, as
shown in the right panel.}
\end{figure}

However, this setup is challenging to generalize to higher-dimensional
black holes. The primary difficulty lies in the lack of a concrete
method for computing the entanglement entropy of a CFT$_{D-1}$ in
higher dimensions. Even if such results were known, it would remain
unclear how to construct the corresponding minimal surface $\gamma_{A}$
in the bulk such that it wraps around the higher-dimensional black
hole's event horizon in a manner analogous to the three-dimensional
case. To solve this issue, let us revisit the BTZ black hole from
a different perspective. Consider the maximally extended Penrose diagram
of the BTZ black hole, which features two asymptotic boundaries, each
of which hosts a copy of the dual CFT$_{2}$ \cite{Maldacena:2001kr}.
These two CFTs on asymptotic boundaries are known as the Thermofield
double (TFD) sate. Each point in this diagram represents a spatial
$S^{1}$ circle, as shown in the left panel of figure (\ref{fig:BTZ_EE2}).
From the viewpoint of the boundary CFT$_{2}$, we can compute the
entanglement entropy $S_{\mathrm{vN}}\left(A:B\right)$ between regions
$A$ and $B$ located on the two separate asymptotic boundaries, following
recent developments in the study of disconnected-boundary entanglement
\cite{Cardy:2016fqc,Jiang:2024ijx,Jiang:2024xqz}. Using the RT formula,
we obtain the corresponding geodesic $\gamma\left(A:B\right)$, which
coincides with the event horizon of the BTZ black hole, as illustrated
in the right panel of (\ref{fig:BTZ_EE2}) (Note that the event horizons
of the BTZ black hole and the wormhole are identical). In other words,
since the holographic dual of the thermofield double (TFD) state and
the BTZ black hole share the same Penrose diagram, and both the entanglement
entropy and black hole entropy correspond to areas associated with
specific points on this diagram, one can use entanglement entropy
to probe black hole entropy whenever the corresponding points coincide.
Therefore, the entanglement entropy $S_{\mathrm{vN}}\left(A:B\right)$
in a two-boundary configuration of CFT$_{2}$ offers a viable approach
for recovering the Bekenstein-Hawking entropy of the BTZ black hole.

\begin{figure}[h]
\begin{centering}
\includegraphics[scale=0.35]{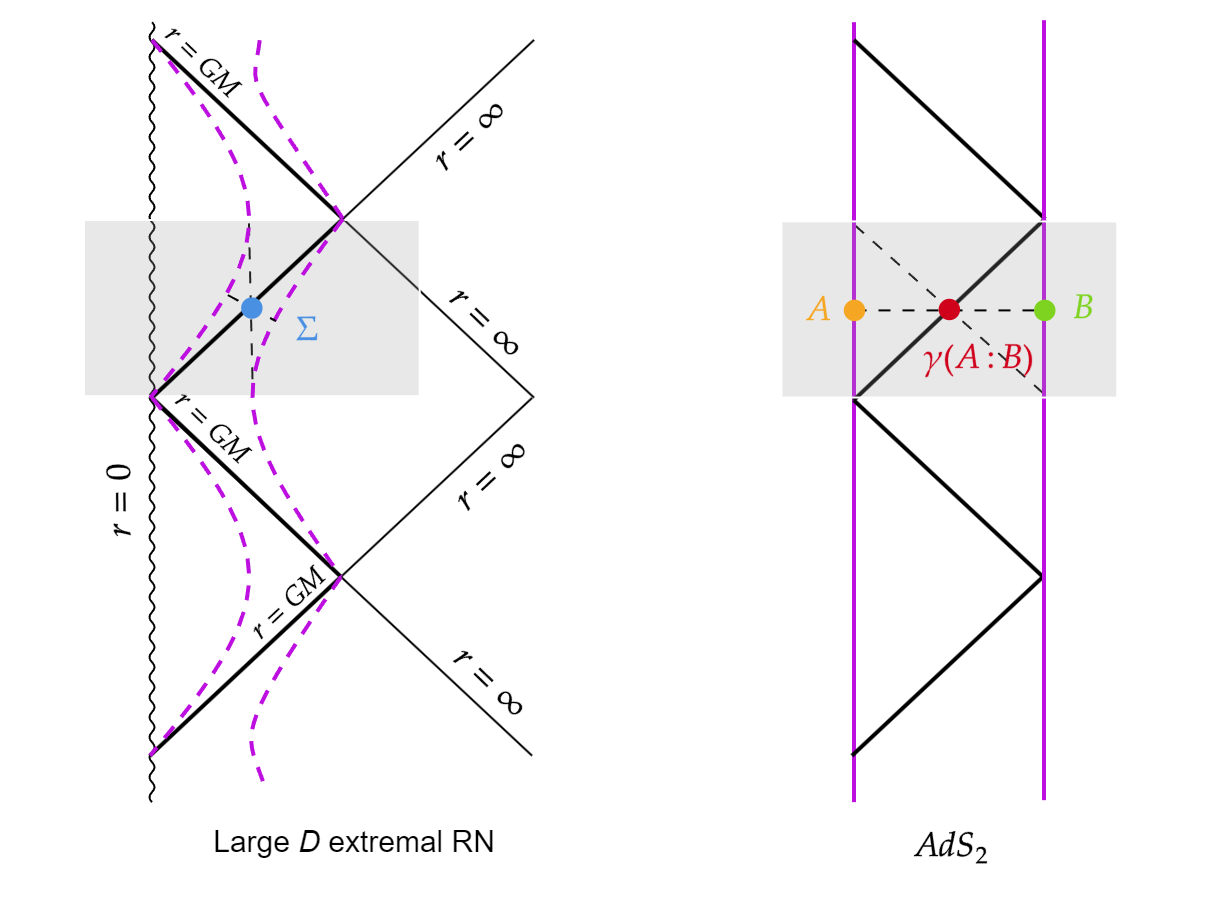}
\par\end{centering}
\caption{\label{fig:RNADS} The shaded square regions on both sides denote
equivalent regions of interest. The left-hand side of the figure shows
the Penrose diagram of a $D$-dimensional extremal RN black hole.
The strip enclosed by dashed purple lines represents its near-horizon
region, which takes the form AdS$_{2}$$\times S^{D-2}$. Each point
along the bold black line corresponds to the same area of the event
horizon. The Bekenstein--Hawking entropy associated with the black
hole is entirely determined by the AdS$_{2}$ geometry, captured by
the blue dot labeled $\mathrm{Area}\left({\color{blue}\Sigma}\right)$.
In the entropy, the higher-dimensional spherical part $S^{D-2}$ of
the spacetime metric is absorbed into the $D$-dimensional Newton\textquoteright s
constant $G_{N}^{\left(D\right)}$, effectively reducing it to a two-dimensional
Newton\textquoteright s constant, denoted $G_{N}^{\left(2\right)}$.
The right-hand side of the figure illustrates the emergent AdS$_{2}$
spacetime derived from the entanglement entropy of a TFD state. In
this setup, the RT prescription yields the area associated with the
red dot $\mathrm{Area}\left({\color{red}\gamma\left(A:B\right)}\right)$.
Since the two purple strip regions represent the same near-horizon
geometry, the areas can be naturally identified: $\mathrm{Area}\left({\color{blue}\Sigma}\right)=\mathrm{Area}\left({\color{red}\gamma\left(A:B\right)}\right)$.
This observation implies that the $D$-dimensional Bekenstein--Hawking
entropy of the extremal black hole can be recovered from the entanglement
entropy of CQM$_{1}$. }
\end{figure}

Now, we aim to generalize this method to arbitrary higher-dimensional
black holes. The key insight lies in the near-horizon region of near-extremal
black holes, as all known extremal black holes exhibit an AdS$_{2}$
factor in their near-horizon geometry \cite{Sen:2008vm}. Based on
this observation, and motivated by the AdS$_{2}$/CFT$_{1}$ correspondence,
it is natural to investigate how the black hole entropy is related
to the entanglement entropy in CFT$_{1}$. Moreover, it is important
to note that the AdS$_{2}$ geometry itself possesses a nonzero entropy,
as it arises in the near-horizon limit of large $D$ near-extremal
black holes \cite{Azeyanagi:2007bj}. In this paper, we focus on a
concrete example: the large $D$ Reissner--Nordström (RN) black hole
of Einstein-Maxwell theory. The correspondence between large $D$
black holes and low-dimensional string theory was first observed in
ref. \cite{Emparan:2013xia}. The near-horizon geometry of the extremal
RN black hole in the large $D$ limit is AdS$_{2}$, which is the
solution of two-dimensional heterotic string effective action. This
solution allows us to compute the entanglement entropy between two
CFT$_{1}$ (or more precisely, conformal quantum mechanics, CQM$_{1}$)
theories defined on the boundaries of AdS$_{2}$. Following the same
reasoning as in the BTZ black hole case, this entanglement entropy
corresponds to a minimal surface $\gamma\left(A:B\right)$, which
in turn captures the area of the event horizon of the large $D$ RN
black hole, as shown in figure (\ref{fig:RNADS}). In the entropy,
the remaining spherical part of the metric will be absorbed into the
$D$-dimensional Newton's constant $G_{N}^{\left(D\right)}$, effectively
reducing it to a two-dimensional quantity, denoted as $G_{N}^{\left(2\right)}$.
Thus, it becomes possible to compute the higher-dimensional Bekenstein--Hawking
entropy from the entanglement entropy of CQM$_{1}$. In the large
$D$ limit, the relation between black hole entropy and entanglement
entropy becomes particularly transparent: black hole entropy arises
entirely from the entanglement entropy between the two CFTs living
on either side of the event horizon. In other words, entanglement
entropy weaves together the structure of black hole entropy.

Several related studies are worth mentioning here. In \cite{Guo:2015swu},
the authors investigate the large $D$ limit of Reissner--Nordström--AdS
(RN-AdS) black holes in both extremal and non-extremal cases. They
show that the near-horizon geometry of the extremal case becomes AdS$_{2}$,
and the Bekenstein--Hawking entropy is computed using the Cardy formula.
Another relevant work examines the island formula in the context of
large $D$ RN-AdS black holes \cite{Tong:2023nvi}. In \cite{Sybesma:2022nby},
the large $D$ limit of Lifshitz black holes is studied. The authors
find that the near-horizon and near-extremal regions are effectively
captured by two-dimensional gravity theories, including the Callan--Giddings--Harvey--Strominger
(CGHS) and Jackiw--Teitelboim (JT) models.

This paper is organized as follows. In Section 2, we review the entanglement
entropy of finite regions in the TFD state. This result can be applied
to extract the Bekenstein--Hawking entropy of the BTZ black hole
and the $D1$-$D5$ black hole in type IIB string theory. In Section
3, we generalize the method to the large $D$ RN black hole. As expected,
the Bekenstein--Hawking entropy exactly matches the entanglement
entropy of CQM$_{1}$. The final section contains our conclusions
and discussions.

\section{Entanglement entropy of the TFD state}

In this section, we briefly review the computation of entanglement
entropy between two disconnected CFTs in the thermofield double (TFD)
formalism. This approach was originally formulated in \cite{Cardy:2016fqc}
and further developed in \cite{Jiang:2024ijx,Jiang:2024xqz}. For
a comprehensive review of the TFD state and its associated Euclidean
path integral construction, see \cite{Hartman:2015,Callebaut:2023fnf}.

\subsection{BTZ black hole}

We consider the total Hilbert space as the tensor product of two identical
CFT Hilbert spaces:

\begin{equation}
\mathcal{H}_{total}=\mathcal{H}_{1}\otimes\mathcal{H}_{2},
\end{equation}

\noindent where each energy eigenstate satisfies $H\left|n\right\rangle =E_{n}\left|n\right\rangle $.
The TFD state, a pure entangled state in this doubled system, is defined
by

\begin{equation}
\left|TFD\right\rangle =\frac{1}{\sqrt{Z\left(\beta\right)}}\underset{n}{\sum}e^{-\beta E_{n}/2}\left|E_{n}\right\rangle _{1}\otimes\left|E_{n}\right\rangle _{2},\qquad Z\left(\beta\right)=\underset{n}{\sum}e^{-\beta E_{n}},
\end{equation}

\noindent where $T=1/\beta$ denotes the temperature. The corresponding
density matrix is given by

\begin{equation}
\rho_{total}=\left|TFD\right\rangle \left\langle TFD\right|.
\end{equation}

\begin{figure}[H]
\begin{centering}
\includegraphics[scale=0.3]{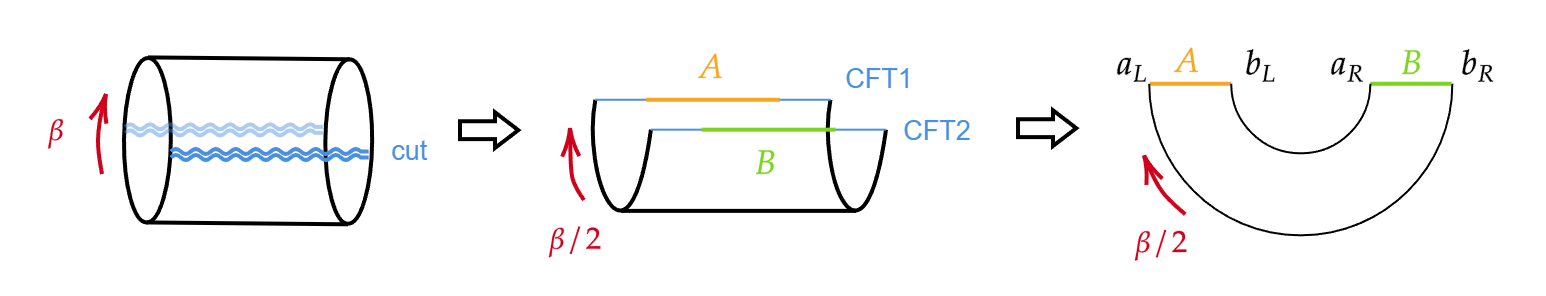}
\par\end{centering}
\caption{\label{fig:TFD} This figure illustrates the preparation of the TFD
state. The first step is to consider the thermal partition function
at temperature $T=1/\beta$. A quantum field theory (QFT) at finite
temperature can be formulated as a Euclidean path integral with imaginary
time periodicity, where $t\sim t+i\beta$. For a two-dimensional theory
on the line, the corresponding Euclidean geometry is an infinite cylinder
with periodicity $\beta$ in the Euclidean time direction. The second
step involves cutting this cylinder along a time interval of length
$\beta/2$, effectively dividing it into two halves. Two copies of
the CFT are then defined on the two resulting blue boundaries, representing
the open cuts. Upon specifying entangling regions $A$ and $B$ on
each side, the relevant path integral region corresponds to a half-annulus
geometry.}
\end{figure}

\noindent The entropy of system 1 is precisely the entanglement entropy
between the two CFT copies, which can be given by:

\begin{equation}
S_{1}=-tr\rho_{1}\ln\rho_{1},
\end{equation}

\noindent where the reduced density matrix is $\rho_{1}=tr_{2}\rho_{total}=e^{-\beta H_{1}}$.
Furthermore, the TFD state can be constructed by \textquotedblleft cutting\textquotedblright{}
the thermal partition function into two halves, as illustrated in
figure (\ref{fig:TFD}). Once the TFD state is prepared, one can specify
the entangling regions $A$ and $B$ on each respective boundary and
compute the corresponding entanglement entropy. 

\begin{figure}[H]
\begin{centering}
\includegraphics[scale=0.35]{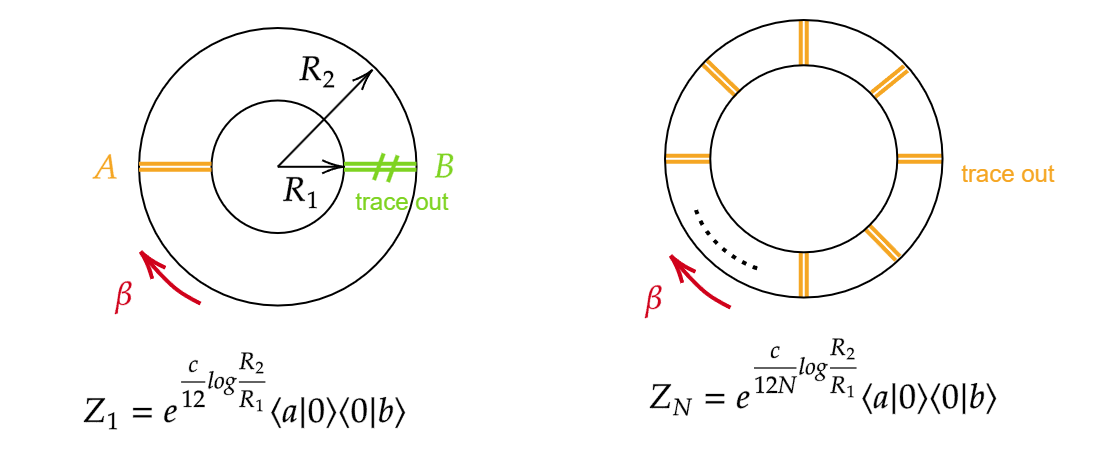}
\par\end{centering}
\caption{\label{fig:replica}The left-hand side of the figure depicts a standard
annulus geometry, whose corresponding partition function is denoted
by $Z_{1}$. The width of the annulus is given by $W=\log\frac{R_{2}}{R_{1}}$,
where $R_{1}$ and $R2$ are the inner and outer radii, respectively.
After applying the replica trick to compute the entanglement entropy,
the geometry is modified to a replicated annulus shown on the right-hand
side. This replicated geometry has a reduced width $W_{N}=\frac{W}{N}$,
and the associated partition function is denoted by $Z_{N}$.}
\end{figure}

The corresponding entanglement entropy in the annular CFT can be computed
using the modular Hamiltonian method, and is given by the replica
trick:

\begin{equation}
S_{\mathrm{vN}}\left(A:B\right)=\underset{N\rightarrow1}{\lim}\left[\frac{1}{1-N}\log tr_{A}\rho_{A}^{N}\right]=\underset{N\rightarrow1}{\lim}\left[\frac{1}{1-N}\log\frac{Z_{N}}{Z_{1}^{N}}\right],
\end{equation}

\noindent where $Z_{1}$ denotes the ordinary annulus partition function
with width $W=\log\frac{R_{2}}{R_{1}}$, and $Z_{N}$ represents the
partition function on the replicated manifold, which is conformally
equivalent to an annulus of width $W_{N}=\frac{W}{N}$, as illustrated
in figure (\ref{fig:replica}).

\noindent Using this setup, the entanglement entropy becomes

\begin{equation}
S_{\mathrm{vN}}\left(A:B\right)=\frac{c}{6}\log\left(1+2\eta^{-1}+2\sqrt{\eta^{-1}\left(\eta^{-1}+1\right)}\right),
\end{equation}
where we neglect boundary contributions, which do not affect the universal
part of the result. The cross-ratio $\eta$ is defined as

\begin{equation}
\eta=\frac{\cosh\left(\frac{\pi\left(a_{L}-a_{R}\right)}{\beta}\right)\cosh\left(\frac{\pi\left(b_{L}-b_{R}\right)}{\beta}\right)}{\sinh\left(\frac{\pi\left(a_{L}-b_{L}\right)}{\beta}\right)\sinh\left(\frac{\pi\left(a_{R}-b_{R}\right)}{\beta}\right)}.
\end{equation}

\noindent By applying the coordinate transformation described in \cite{Jiang:2024xqz},
and specializing to the case $a_{L}=a_{R}=a$ and $b_{L}=b_{R}=b$,
the expression simplifies to

\begin{equation}
S_{\mathrm{vN}}\left(A:B\right)=\frac{\pi c}{3\beta}\left|a-b\right|.\label{eq:TFD EE}
\end{equation}

\noindent Using the AdS/CFT dictionary: $c=3R_{AdS}/2G_{N}^{\left(3\right)}$,
$\frac{\left|a-b\right|}{\beta}=\frac{r_{+}}{R_{AdS}}$, and the Hawking
temperature $\beta=\frac{2\pi R_{AdS}^{2}}{r_{+}}$, we obtain

\begin{equation}
S_{\mathrm{vN}}\left(A:B\right)=\frac{2\pi r_{+}}{4G_{N}^{\left(3\right)}}.
\end{equation}

\begin{figure}[H]
\begin{centering}
\includegraphics[scale=0.3]{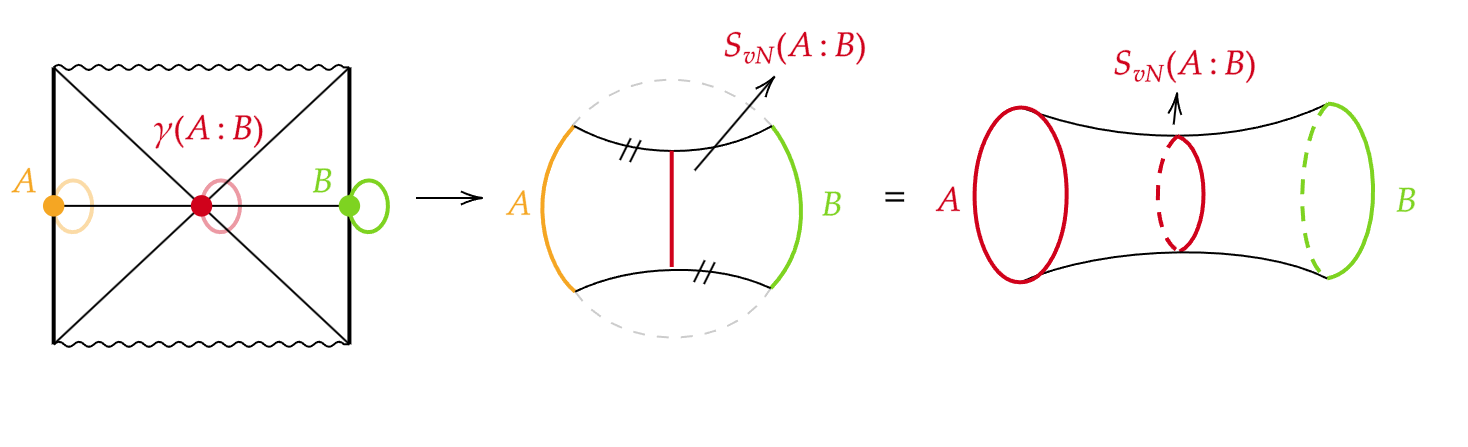}
\par\end{centering}
\caption{\label{fig:BTZ BH}This picture illustrates the spacetime that emerges
from the entanglement entropy of the TFD state. The spacetime geometry
is reconstructed via Synge\textquoteright s world function associated
with the von Neumann entanglement entropy $S_{\mathrm{vN}}\left(A:B\right)$.
In the first panel, each point along the central horizontal line corresponds
to an $S^{1}$ circle. Therefore, this horizontal line can be visualized
as in the second or third panels. The third panel is obtained as a
quotient of the geometry shown in the second panel.}
\end{figure}

\noindent This value precisely reproduces the area of the red point
(entangling surface) according to the RT formula in figure (\ref{fig:BTZ BH}),
which coincides with the area of the BTZ black hole event horizon
$\Sigma$, since the entanglement entropy and black hole entropy correspond
to the same surface in the shared Penrose diagram of the TFD state
and BTZ geometry. Therefore, we conclude:

\begin{equation}
S_{\mathrm{vN}}\left(A:B\right)=S_{BH},
\end{equation}

\noindent establishing a direct identification between entanglement
entropy in the TFD state and the Bekenstein--Hawking entropy of the
BTZ black hole. 

\subsection{$D1$-$D5$ black hole of type IIB string theory}

A quick example is the $D1$-$D5$ black hole, which arises as a solution
of ten-dimensional type IIB string theory with $Q_{1}$ D$1$-branes
and $Q_{5}$ D$5$-branes wrapped on a compact internal space $T^{4}\times S^{1}$.
Here, the four-torus $T^{4}$ has volume $\left(2\pi\right)^{4}v\alpha^{\prime}$,
and the circle $S^{1}$ has radius $R$ \cite{Maldacena:1998bw}.
The Kaluza-Klein momentum along $S^{1}$ is quantized and labeled
by $N$. The near-horizon limit is obtained by taking the string length
squared $\alpha^{\prime}\rightarrow0$, leading to the following geometry:

\begin{equation}
\frac{dS^{2}}{\alpha^{\prime}}=\left[\frac{U^{2}}{\ell^{2}}\left(-dt^{2}+\left(dx^{5}\right)^{2}\right)+\frac{\ell^{2}}{U^{2}-U_{0}^{2}}dU^{2}\right]+\frac{U_{0}^{2}}{\ell^{2}}\left(\cosh\sigma dt+\sinh\sigma dx^{5}\right)^{2}+\ell^{2}\left(d\Omega_{3}\right)^{2}+\sqrt{\frac{Q_{1}}{vQ_{5}}}dx_{i}^{2},\label{eq:D5 BH}
\end{equation}

\noindent where $\ell^{2}=g_{6}\sqrt{Q_{1}Q_{5}}$, $g_{6}$ is the
six-dimensional string coupling, and $x^{5}$ is the coordinate along
$S^{1}$. Note $Q_{1}Q_{5}$ also determines the central charge of
the dual two-dimensional CFT via the Kac-Moody superconformal algebra.
It is evident that the coordinates $\left(x^{5},U,t\right)$ span
a locally AdS$_{3}$ geometry. Therefore, the full near-horizon geometry
(\ref{eq:D5 BH}) can be interpreted as 

\begin{equation}
\mathrm{BTZ}_{3}\times S^{3}\times T^{4}.
\end{equation}

\noindent The corresponding Bekenstein--Hawking entropy is given
by the Cardy formula for the dual CFT:

\begin{equation}
S_{BH}=2\pi\sqrt{\frac{cL_{0}}{6}}=2\pi\sqrt{Q_{1}Q_{5}N}.\label{eq:5D BH entropy}
\end{equation}

To probe this result using entanglement entropy, we also compute the
entanglement entropy between the two CFTs defined on the asymptotic
boundaries. Since the near-horizon region of the $D1$-$D5$ black
hole is $\mathrm{BTZ}_{3}$, our previous result for the entanglement
entropy of the TFD (\ref{eq:TFD EE}) is applicable here. However,
given the specific form of the near-horizon metric (\ref{eq:D5 BH}),
the central charge of the dual CFT must include contributions from
both bosonic and fermionic degrees of freedom. This yields \cite{Strominger:1996sh,Callan:1996dv}:

\begin{equation}
c=6Q_{1}Q_{5}.
\end{equation}

\noindent Furthermore, from the near-horizon metric (\ref{eq:D5 BH}),
the AdS radius can be identified as

\begin{equation}
R_{AdS}^{2}=\ell^{2}.
\end{equation}

\noindent In the extremal limit, the black hole radius $r_{+}$ is
related to the microscopic parameters by \cite{Maldacena:1998bw}:

\begin{equation}
\frac{r_{+}^{2}}{R_{AdS}^{2}}=\frac{N}{Q_{1}Q_{5}}.
\end{equation}

\noindent Substituting this into our general entanglement entropy
result (\ref{eq:TFD EE}), and using the identification $\frac{\left|a-b\right|}{\beta}=\frac{r_{+}}{R_{AdS}}$,
we find
\begin{equation}
S_{\mathrm{vN}}\left(A:B\right)=\frac{\pi c}{3\beta}\left|a-b\right|=2\pi\sqrt{Q_{1}Q_{5}N}.\label{eq:5D EE}
\end{equation}

\noindent which precisely reproduces the Bekenstein--Hawking entropy
given in (\ref{eq:5D BH entropy}). This agreement confirms that our
method, based on computing entanglement entropy from the dual CFT,
correctly captures the Bekenstein--Hawking entropy of the $D1$-$D5$
black hole, whose near-horizon geometry is BTZ$_{3}$. 

Moreover, it is worth noting that the near horizon limit of near extremal
BTZ is AdS$_{2}$$\times S^{1}$. This implies that the corresponding
TFD state in AdS$_{2}$/CFT$_{1}$ can also be used to compute the
entanglement entropy. In this case, one starts with a rotating BTZ
black hole, which modifies the CFT defined on the asymptotic boundary.
The TFD state must then be replaced by \cite{Maldacena:2001kr}:

\begin{equation}
\left|TFD\right\rangle =\frac{1}{\sqrt{Z_{0}}}\underset{n_{L},n_{R}}{\sum}e^{-\frac{\beta_{L}L_{0}}{2}-\frac{\beta_{R}\bar{L}_{0}}{2}}\left|n_{L},n_{R}\right\rangle _{1}\otimes\left|n_{L},n_{R}\right\rangle _{2},
\end{equation}

\noindent where $Z_{0}$ is the partition function of the CFT$_{2}$.
Consequently, the corresponding entanglement entropy differs from
the non-rotating case. This entropy has been computed in \cite{Azeyanagi:2007bj}
by taking the near-extremal limit---where $\beta_{L}$ remains finite
and $\beta_{R}\rightarrow\infty$ ---and the near-horizon limit $r\rightarrow r_{+}$.
In this setup, states labeled by $L_{0}=N$ are denoted as $\left|k\right\rangle $,
where $k=1,2,\ldots,d\left(N\right)$. The degeneracy $d\left(N\right)$
is large and grows as:

\begin{equation}
d\left(N\right)\sim e^{2\pi\sqrt{Q_{1}Q_{5}N}}.
\end{equation}

\noindent The TFD state in the near-horizon, near-extremal limit can
then be rewritten as

\begin{equation}
\left|TFD\right\rangle =\frac{1}{\sqrt{d\left(N\right)}}\underset{n}{\sum}\underset{k=1}{\overset{d\left(N\right)}{\sum}}e^{-\frac{\beta E_{n}}{2}}\left|k_{L},n_{R}\right\rangle _{\mathrm{CQM}1}\otimes\left|k_{L},n_{R}\right\rangle _{\mathrm{CQM}2},
\end{equation}

\noindent where $E_{n}=\left\langle n\left|\bar{L}_{0}\right|n\right\rangle $.
In the zero temperature limit $\beta=\infty$, there is only the single
ground state $\left|0\right\rangle $ in the the right-moving sector,
and the state reduces to $\left|k\right\rangle _{\mathrm{CQM}1}=\left|k_{L},0\right\rangle $.
The reduced density matrix of CQM1, obtained by tracing out CQM2 ,
is

\begin{equation}
\rho_{1}=\frac{1}{\sqrt{d\left(N\right)}}\underset{k=1}{\overset{d\left(N\right)}{\sum}}\left|k\right\rangle \left\langle k\right|_{\mathrm{CQM}1}.
\end{equation}

\noindent The entanglement entropy then becomes

\begin{equation}
S_{1}=-tr\rho_{1}\ln\rho_{1}=\log d\left(N\right)=2\pi\sqrt{Q_{1}Q_{5}N},
\end{equation}

\noindent which exactly matches our previous result (\ref{eq:5D EE}).
This equivalence demonstrates that our approach not only captures
the Bekenstein--Hawking entropy but also provides a microscopic counting
of BPS states of the black hole.

\section{Large $D$ black hole entropy and CQM$_{1}$ entanglement entropy}

\begin{figure}[h]
\begin{centering}
\includegraphics[scale=0.4]{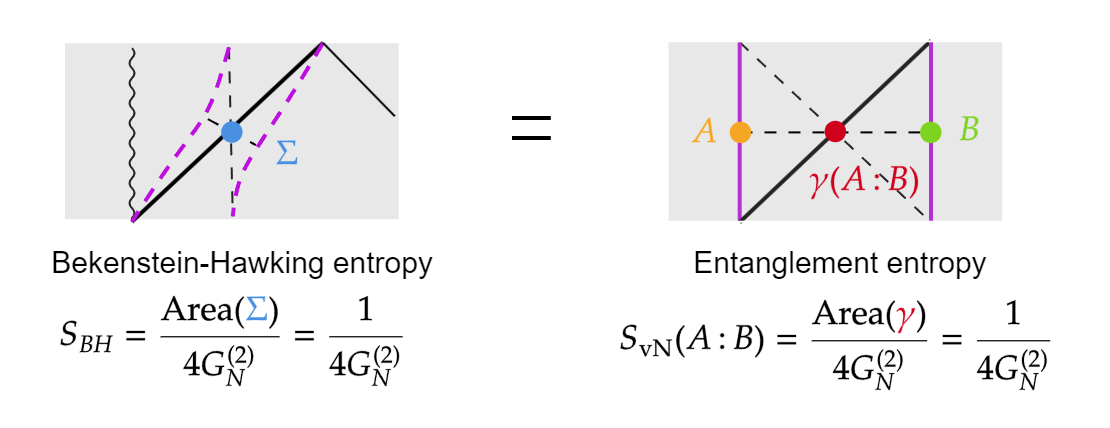}
\par\end{centering}
\caption{\label{fig:equ}This figure highlights the shaded regions of interest
from figure (\ref{fig:RNADS}). The purple strip regions on both sides
share the same Penrose diagram and describe identical near-horizon
geometries. In this section, we demonstrate that the Bekenstein--Hawking
entropy of the RN black hole in the large $D$ limit, shown on the
left-hand side, precisely matches the entanglement entropy of the
CQM$_{1}$ on the right-hand side.}
\end{figure}

In this section, we demonstrate that the entanglement entropy of a
one-dimensional CQM$_{1}$ precisely matches the Bekenstein--Hawking
entropy of a RN black hole in the large $D$ limit. Specifically,
we begin by analyzing the RN black hole in the large $D$ regime,
where the near-horizon geometry simplifies and becomes effectively
described by a two-dimensional charged black hole---a solution arising
from the heterotic string effective action. In the extremal limit,
this near-horizon geometry further reduces to AdS$_{2}$. Simultaneously,
we examine the Bekenstein--Hawking entropy of the RN black hole under
the same limiting procedure: large $D$ and extremality. In this limit,
the entropy becomes a purely two-dimensional quantity, as the area
of the event horizon shrinks to a point (represented by the blue dot
in the left panel of figure (\ref{fig:equ})). On the other hand,
we compute the entanglement entropy for two disconnected CQM$_{1}$
systems using both field-theoretic and holographic techniques. The
results obtained from these independent methods are in complete agreement
and yield a minimal surface located at the midpoint of the Penrose
diagram (represented by the red dot in the right panel of figure (\ref{fig:equ})).
As expected, this entanglement entropy exactly reproduces the Bekenstein--Hawking
entropy of the extremal RN black hole in the large $D$ limit, thereby
establishing a precise correspondence between the two.

\subsection{Bekenstein-Hawking entropy of RN black hole at large $D$ }

In this subsection, we initiate the computation of the large $D$
limit of the RN black hole and its corresponding Bekenstein--Hawking
entropy. Given the one-to-one correspondence between black hole solutions
and gravitational actions, the large $D$ behavior can be analyzed
from two complementary perspectives: the metric description and the
action-based formulation. Accordingly, our strategy proceeds along
the following route: 
\begin{center}
\includegraphics[scale=0.3]{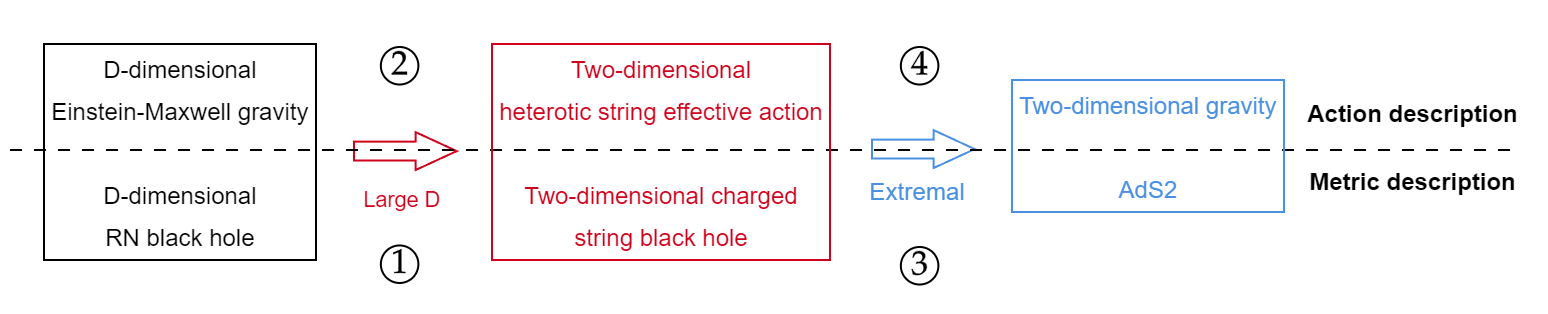}
\par\end{center}

\subsubsection{RN black hole at large $D$ }

\noindent \textbf{Metric description:}
\begin{center}
\includegraphics[scale=0.3]{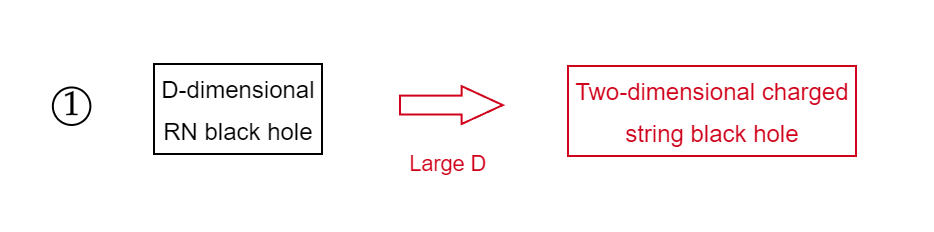}
\par\end{center}

\noindent Let us begin with the first step $\textcircled{1}$, where
we aim to demonstrate that the large $D$ limit of the RN black hole
reduces to the two-dimensional charged black hole solution in string
theory. We first recall the $D$-dimensional RN black hole solution
in Einstein--Maxwell theory \cite{Myers:1986un}:

\begin{equation}
ds^{2}=-f\left(r\right)dt^{2}+\frac{dr^{2}}{f\left(r\right)}+r^{2}d\Omega_{n+1}^{2}\label{eq:metric 1}
\end{equation}

\noindent where $D=n+3$, and

\begin{equation}
f\left(r\right)=1-\frac{2m}{r^{n}}+\frac{q^{2}}{r^{2n}},\qquad A_{t}=\sqrt{\frac{n+1}{2n}}\left(C-\frac{q}{r^{n}}\right)dt,
\end{equation}

\noindent with $C$ an arbitrary constant. The parameters $m$ and
$q$ are related to the ADM mass $M$ and the physical electric charge
$Q$ via:

\begin{equation}
m=\frac{8\pi}{\Omega_{n+1}\left(n+1\right)}M,\qquad q^{2}=\frac{2}{\left(n+1\right)n}Q^{2},\qquad\Omega_{n+1}=\frac{2\pi^{\left(n+2\right)/2}}{\Gamma\left(\frac{n+2}{2}\right)}.
\end{equation}

\noindent The inner and outer horizons $r_{\pm}$ are determined by
the roots of:

\begin{equation}
1-\frac{r_{0}^{n}}{r^{n}}+\frac{q^{2}}{r^{2n}}=0,
\end{equation}

\noindent where we define $r_{0}^{n}\equiv2m$, so that $\mathrm{dim}\left[r_{0}\right]=\mathrm{Length}$.
Accordingly, $\mathrm{dim}\left[m\right]=\mathrm{dim}\left[q\right]=\mathrm{Length}^{n}$.
The horizon locations are given by:

\begin{equation}
r_{\pm}^{n}=\frac{1}{2}r_{0}^{n}\pm\sqrt{\frac{1}{4}r_{0}^{2n}-q^{2}}.
\end{equation}

\noindent In the extremal case, the condition $q^{2}=\frac{1}{4}r_{0}^{2n}$
yields a degenerate horizon at:

\begin{equation}
r_{h}^{n}\equiv r_{\pm}^{n}=\frac{1}{2}r_{0}^{n}=m.\label{eq:finite}
\end{equation}

\noindent To study the near-horizon geometry, we introduce the dimensionless
coordinate:

\begin{equation}
\mathrm{R}=\left(\frac{r}{r_{0}}\right)^{n},
\end{equation}

\noindent Before proceeding, it is worth emphasizing that the notion
of the event horizon differs significantly between finite and large
$D$. To see this, we express $f\left(r\right)$ in terms of $\mathrm{R}$:

\begin{equation}
f\left(\mathrm{R}\right)=1-\frac{1}{\mathrm{R}}+\frac{q^{2}}{\left(2m\right)^{2}}\frac{1}{\mathrm{R}^{2}}.
\end{equation}

\noindent It follows that:

\begin{equation}
\mathrm{R}_{\pm}=\left(\frac{r_{\pm}}{r_{0}}\right)^{n}\Rightarrow r_{\pm}=r_{0}\mathrm{R}_{\pm}^{\frac{1}{n}}\rightarrow r_{\pm}=r_{0},\label{eq:infinite}
\end{equation}

\noindent in the large $D$ limit. This result clearly contrasts with
the finite $D$ case (\ref{eq:finite}). Let us now rewrite the metric
in terms of the coordinate $\mathrm{R}$:

\begin{eqnarray}
ds^{2} & = & -\left(1-\frac{1}{\mathrm{R}}+\frac{q^{2}}{\left(2m\right)^{2}}\frac{1}{\mathrm{R}^{2}}\right)dt^{2}+\left(\frac{r_{0}}{n}\right)^{2}\frac{\mathrm{R}^{\frac{2\left(1-n\right)}{n}}d\mathrm{R}^{2}}{\left(1-\frac{1}{\mathrm{R}}+\frac{q^{2}}{\left(2m\right)^{2}}\frac{1}{\mathrm{R}^{2}}\right)}+r_{0}^{2}\mathrm{R}^{\frac{2}{n}}d\Omega_{n+1}^{2},\nonumber \\
A_{t} & = & \sqrt{\frac{n+1}{2n}}\left(C-\frac{q}{r_{0}^{n}\mathrm{R}}\right)dt.
\end{eqnarray}

\noindent Note that we have not assumed $\mathrm{R}=\mathrm{R}_{\pm}$,
so this result holds for both extremal and non-extremal cases. In
the large $D$ limit (i.e., $\ln\mathrm{R}\ll n$), the metric simplifies
to:

\begin{eqnarray}
ds^{2} & = & -\left(1-\frac{1}{\mathrm{R}}+\frac{q^{2}}{\left(2m\right)^{2}}\frac{1}{\mathrm{R}^{2}}\right)dt^{2}+\left(\frac{r_{0}}{n}\right)^{2}\frac{d\mathrm{R}^{2}}{\mathrm{R}^{2}\left(1-\frac{1}{\mathrm{R}}+\frac{q^{2}}{\left(2m\right)^{2}}\frac{1}{\mathrm{R}^{2}}\right)}+r_{0}^{2}d\Omega_{n+1}^{2},\nonumber \\
A_{t} & = & \sqrt{\frac{1}{2}}\left(C-\frac{q}{\mathrm{R}}\frac{1}{r_{0}^{n}}\right)dt.
\end{eqnarray}

\noindent To restore the physical radial dimension, we define:

\begin{equation}
\mathrm{R}=\frac{\bar{\mathrm{R}}}{2m},\qquad2\lambda=\frac{n}{r_{0}}.\label{eq:dim coordinate trans}
\end{equation}

\noindent Since $\bar{\mathrm{R}}=2m\frac{r^{n}}{r_{0}^{n}}=r^{n}$,
this transformation connects $\bar{\mathrm{R}}$ with the $n$-th
power of the physical radius $r$. This form precisely matches the
two-dimensional charged black hole solution of heterotic string theory
\cite{McGuigan:1991qp,Giveon:2005jv}:

\begin{eqnarray}
ds^{2} & = & -\left(1-\frac{2m}{\bar{\mathrm{R}}}+\frac{q^{2}}{\bar{\mathrm{R}}^{2}}\right)dt^{2}+\frac{d\bar{\mathrm{R}}^{2}}{\mathrm{\left(2\lambda\right)}^{2}\bar{\mathrm{R}}^{2}\left(1-\frac{2m}{\bar{\mathrm{R}}}+\frac{q^{2}}{\bar{\mathrm{R}}^{2}}\right)}+r_{0}^{2}d\Omega_{n+1}^{2},\nonumber \\
A_{t} & = & \sqrt{\frac{1}{2}}\left(\frac{q}{\bar{\mathrm{R}}_{+}}-\frac{q}{\bar{\mathrm{R}}}\right)dt.
\end{eqnarray}

\noindent with field strength $F^{2}=-\frac{4\lambda^{2}q^{2}}{\bar{\mathrm{R}}^{2}}$.
The gauge potential is fixed by requiring $A_{t}\left(\bar{\mathrm{R}}_{+}\right)=0$.
The dilaton field in this background is given by $\phi\left(\bar{\mathrm{R}}\right)=-\frac{1}{2}\ln\frac{\bar{\mathrm{R}}}{2m}$,
which can be derived from the corresponding low-energy effective action. 

\vspace*{2.0ex}

\noindent \textbf{Action description:}
\begin{center}
\includegraphics[scale=0.3]{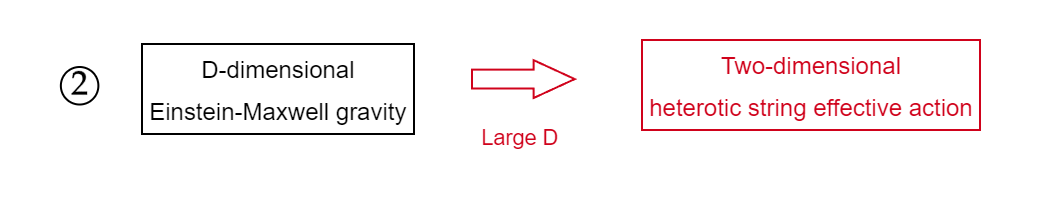}
\par\end{center}

\noindent Now, we consider step $\textcircled{2}$, which involves
the action description corresponding to the black hole solution obtained
in step $\textcircled{1}$. We begin with the $D$-dimensional Einstein--Maxwell
theory, whose action is given by: 

\begin{equation}
I_{\mathrm{EM}}=\frac{1}{16\pi G_{N}^{\left(D\right)}}\int d^{D}x\sqrt{-g}\left(R-F_{\mu\nu}F^{\mu\nu}\right)
\end{equation}

\noindent To introduce a dilaton field $\phi(x^{\mu})$, we perform
a dimensional reduction on a $\left(n+1\right)$-dimensional sphere,
with the following ansatz for the metric: 

\begin{equation}
ds^{2}=\underset{2\;\mathrm{dimensions}}{\underbrace{\mathbb{G}_{\mu\nu}\left(x^{\mu}\right)dx^{\mu}dx^{\nu}}}+\underset{n+1\;\mathrm{dimensional\;sphere}}{\underbrace{r_{0}^{2}e^{-4\phi\left(x^{\mu}\right)/\left(n+1\right)}d\Omega_{n+1}^{2}}},\label{eq:metric 2}
\end{equation}

\noindent where $\mathbb{G}_{\mu\nu}$ is the two-dimensional metric,
and $d\Omega_{n+1}^{2}$ is the line element of a unit $\left(n+1\right)$-sphere.
Under this reduction, the Einstein--Maxwell action becomes:

\begin{equation}
I_{\mathrm{EM}}=\frac{\Omega_{n+1}r_{0}^{n+1}}{16\pi G_{N}^{\left(D\right)}}\int d^{2}x\sqrt{-\mathbb{G}}e^{-2\phi}\left(\mathbb{R}+\frac{4n}{n+1}\left(\nabla\phi\right)^{2}+\frac{n\left(n+1\right)}{r_{0}^{2}}e^{-4\phi/\left(n+1\right)}-F_{\mu\nu}F^{\mu\nu}\right),
\end{equation}

\noindent where $\mathbb{R}$ is the Ricci scalar associated with
the 2$D$ metric, and the volume of the unit $\left(n+1\right)$-sphere
is given by $\Omega_{n+1}=2\pi^{\frac{n+2}{2}}/\Gamma\left(\frac{n+2}{2}\right)$.
Taking the large $n$ limit ($n\rightarrow\infty$), the action reduces
to the two-dimensional string effective action:

\begin{equation}
I_{\mathrm{string}}=\frac{1}{16\pi G_{N}^{\left(2\right)}}\int d^{2}x\sqrt{-\mathbb{G}}e^{-2\phi}\left(\mathbb{R}+4\left(\nabla\phi\right)^{2}+4\lambda^{2}-F^{2}\right),
\end{equation}

\noindent where $\lambda=\frac{n}{2r_{0}}$, and the effective two-dimensional
Newton\textquoteright s constant is given by

\begin{equation}
G_{N}^{\left(2\right)}=\underset{n\rightarrow\infty}{\lim}\frac{G_{N}^{\left(D\right)}}{\Omega_{n+1}r_{0}^{n+1}}.\label{eq:Newton}
\end{equation}

\noindent This is precisely the two-dimensional heterotic string effective
action. To determine the dilaton profile, we compare the metrics (\ref{eq:metric 1})
and (\ref{eq:metric 2}), which leads to:

\begin{equation}
\phi\left(\mathrm{R}\right)=-\frac{n+1}{2n}\ln\mathrm{R}\overset{n\rightarrow\infty}{\longrightarrow}-\frac{1}{2}\ln\mathrm{R}.
\end{equation}

\noindent Applying the same coordinate transformation on $\mathrm{R}$
as used in (\ref{eq:dim coordinate trans}), we obtain the dilaton
solution in terms of the rescaled coordinate $\bar{\mathrm{R}}$:

\begin{equation}
\phi\left(\bar{\mathrm{R}}\right)=-\frac{1}{2}\ln\frac{\bar{\mathrm{R}}}{2m}.
\end{equation}

\subsubsection{Extremal limit of RN black hole at large $D$}

\noindent \textbf{Metric description:}
\begin{center}
\includegraphics[scale=0.3]{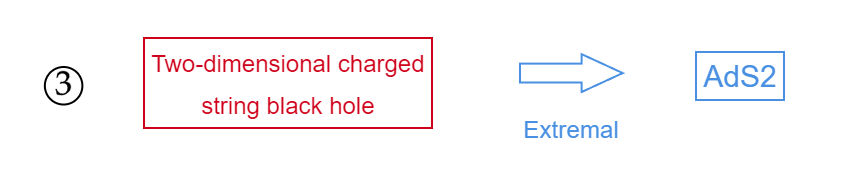}
\par\end{center}

\noindent In this subsection, we proceed to discuss steps $\textcircled{3}$
and $\textcircled{4}$, which involve taking the extremal limit of
the large $D$ black hole and analyzing its associated action. Recall
that the inner and outer horizons of a large $D$ black hole are determined
by the condition: 

\begin{equation}
1-\frac{2m}{\bar{\mathrm{R}}}+\frac{q^{2}}{\bar{\mathrm{R}}^{2}}=0,
\end{equation}

\noindent Solving this equation yields the inner and outer horizons
$\bar{\mathrm{R}}_{\pm}$:

\begin{equation}
\bar{\mathrm{R}}_{\pm}=m\pm\sqrt{m^{2}-q^{2}},
\end{equation}

\noindent or, equivalently,

\begin{equation}
\bar{\mathrm{R}}_{+}+\bar{\mathrm{R}}_{-}=2m,\qquad\bar{\mathrm{R}}_{+}\bar{\mathrm{R}}_{-}=q^{2}.
\end{equation}

\noindent Using this notation, the black hole metric can be rewritten
as:

\begin{equation}
ds^{2}=-\frac{1}{\bar{\mathrm{R}}^{2}}\left(\bar{\mathrm{R}}-\bar{\mathrm{R}}_{+}\right)\left(\bar{\mathrm{R}}-\bar{\mathrm{R}}_{-}\right)dt^{2}+\left(\frac{r_{0}}{n}\right)^{2}\frac{d\bar{\mathrm{R}}^{2}}{\left(\bar{\mathrm{R}}-\bar{\mathrm{R}}_{+}\right)\left(\bar{\mathrm{R}}-\bar{\mathrm{R}}_{-}\right)}+r_{0}^{2}d\Omega_{D-2}^{2},
\end{equation}

\noindent with gauge field and dilaton:

\begin{equation}
A_{t}=\sqrt{\frac{1}{2}}\frac{q}{\bar{\mathrm{R}}_{+}}\left(1-\frac{\bar{\mathrm{R}}_{+}}{\bar{\mathrm{R}}}\right)dt,\qquad\phi\left(\bar{\mathrm{R}}\right)=-\frac{1}{2}\ln\frac{\bar{\mathrm{R}}}{2m}.
\end{equation}

\noindent To study the near-extremal and near-horizon limit, we follow
the coordinate transformations introduced in \cite{Guo:2015swu}:

\begin{equation}
\bar{\mathrm{R}}\rightarrow\bar{\mathrm{R}}_{+}+\epsilon\tilde{\mathrm{R}},\qquad\bar{\mathrm{R}}_{+}-\bar{\mathrm{R}}_{-}=\epsilon,\qquad t\rightarrow\frac{\bar{\mathrm{R}}_{+}}{\epsilon}\frac{r_{0}}{n}\tau.
\end{equation}

\noindent Applying this limit, the metric becomes

\begin{eqnarray}
ds^{2} & = & \left(\frac{r_{0}}{n}\right)^{2}\left[-\left(\tilde{\mathrm{R}}\right)\left(\tilde{\mathrm{R}}+1\right)d\tau^{2}+\frac{d\tilde{\mathrm{R}}^{2}}{\left(\tilde{\mathrm{R}}\right)\left(1+\tilde{\mathrm{R}}\right)}\right]+r_{0}^{2}d\Omega_{D-2}^{2},\nonumber \\
A_{\tau} & = & \sqrt{\frac{1}{2}}\frac{q}{\bar{\mathrm{R}}_{+}}\tilde{\mathrm{R}}\frac{r_{0}}{n}d\tau,\nonumber \\
\phi\left(\bar{\mathrm{R}}\right) & = & -\frac{1}{2}\ln\frac{\bar{\mathrm{R}}_{+}}{2m}.
\end{eqnarray}

\noindent We can also rewrite the metric as:

\begin{equation}
ds^{2}=\left(\frac{r_{0}}{n}\right)^{2}\left[-\left(\left(\tilde{\mathrm{R}}+\frac{1}{2}\right)^{2}-\frac{1}{4}\right)d\tau^{2}+\frac{d\tilde{\mathrm{R}}^{2}}{\left(\left(\tilde{\mathrm{R}}+\frac{1}{2}\right)^{2}-\frac{1}{4}\right)}\right]+r_{0}^{2}d\Omega_{D-2}^{2}.
\end{equation}

\noindent Introducing a new coordinate:

\begin{equation}
\rho=\tilde{\mathrm{R}}+\frac{1}{2},
\end{equation}

\noindent and in the extremal limit, where $q=\bar{\mathrm{R}}_{+}=m$,
the geometry becomes:

\begin{eqnarray}
ds^{2} & = & \left(\frac{r_{0}}{n}\right)^{2}\left[-\left(\rho^{2}-\frac{1}{4}\right)d\tau^{2}+\frac{d\rho^{2}}{\left(\rho^{2}-\frac{1}{4}\right)}\right]+r_{0}^{2}d\Omega_{D-2}^{2},\nonumber \\
A_{\tau} & = & \sqrt{\frac{1}{2}}\frac{r_{0}}{n}\left(\rho-\frac{1}{2}\right)d\tau,\nonumber \\
\phi\left(\rho\right) & = & -\frac{1}{2}\ln\frac{1}{2}.\label{eq:ADS2 solution}
\end{eqnarray}

\noindent The field strength squared becomes: $F^{2}=-\frac{4\lambda^{2}q^{2}}{\bar{\mathrm{R}}^{2}}=-4\lambda^{2}$.
We observe that the complete metric reduces to $\mathrm{AdS_{2}\times S^{D-2}}$,
where the $\left(\tau,\rho\right)$ sector is precisely AdS$_{2}$.
The factor $r_{0}^{2}$ multiplying $d\Omega_{D-2}^{2}$ indicates
that this geometry describes the near-horizon region around $r\rightarrow r_{0}$,
with the transverse sphere remaining of fixed radius. It is important
to emphasize that this AdS$_{2}$ near-horizon geometry arises only
in the extremal limit. For non-extremal black holes, the near-horizon
geometry is approximately $\mathrm{Rindler\times S^{D-2}}$, rather
than $\mathrm{AdS_{2}\times S^{D-2}}$. Thus, the appearance of the
AdS$_{2}$ throat is a direct consequence of extremality. Moreover,
the spherical part $S^{D-2}$ does not affect the conformal boundary
structure of the near-horizon geometry. Near the AdS$_{2}$ boundary,
the conformal factor of the two-dimensional metric diverges, while
the sphere maintains finite size $r_{0}^{2}$. After dimensional reduction,
the entire contribution of the sphere is absorbed into the effective
two-dimensional Newton constant $G_{N}^{\left(2\right)}$. Consequently,
the entropy is controlled by $1/G_{N}^{\left(2\right)}$, and the
original horizon \textquotedblleft area\textquotedblright{} is encoded
in the two-dimensional theory rather than appearing as an explicit
geometric area. In the reduced AdS$_{2}$ description, the event horizon
corresponds to a point on a constant-time slice. In this sense, the
AdS$_{2}$ Penrose diagram can be viewed as a zoomed-in description
of the near-horizon region of the Reissner--Nordström geometry, as
illustrated in figure (\ref{fig:RNADS}). Since AdS$_{2}$ possesses
two asymptotic boundaries, the near-horizon limit isolates a two-boundary
structure from the full RN Penrose diagram. In this limit, the black
hole horizon of the RN geometry is mapped to the Poincaré horizon
of AdS$_{2}$.

\vspace*{2.0ex}

\noindent \textbf{Action description:}
\begin{center}
\includegraphics[scale=0.3]{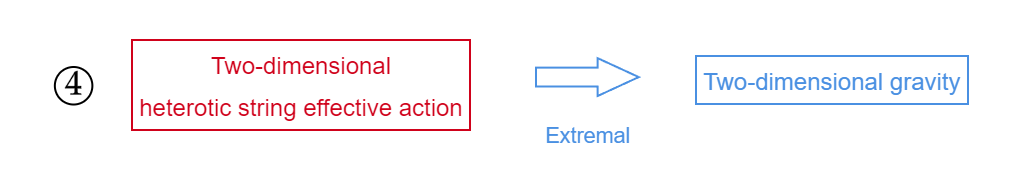}
\par\end{center}

\noindent Finally, let us see how the two-dimensional heterotic string
effective action reduces to two-dimensional gravity in the extremal
limit. Recall the two-dimensional heterotic string effective action: 

\begin{equation}
I_{\mathrm{string}}=\frac{1}{16\pi G_{N}^{\left(2\right)}}\int d^{2}x\sqrt{-\mathbb{G}}e^{-2\phi}\left(\mathbb{R}+4\left(\nabla\phi\right)^{2}+4\lambda^{2}-F^{2}\right).
\end{equation}

\noindent In the extremal limit, we use the constant dilaton solution
$\phi\left(\bar{\mathrm{R}}\right)=-\frac{1}{2}\ln\frac{1}{2}$ given
earlier to integrate out the dilaton. The action then becomes

\begin{equation}
I_{\mathrm{string}}=\frac{1}{16\pi G_{N}^{\left(2\right)}}\int d^{2}xe^{-2\phi}\sqrt{-\mathbb{G}}\left(\mathbb{R}+4\lambda^{2}-F^{2}\right),
\end{equation}
which corresponds to Jackiw--Teitelboim (JT) gravity without the
topological term. The solution to this action is

\begin{equation}
\mathbb{R}=-8\lambda^{2},\qquad F^{2}=-\frac{4\lambda^{2}q^{2}}{\bar{\mathrm{R}}^{2}}=-4\lambda^{2},
\end{equation}

\noindent which agrees with the AdS$_{2}$ solution obtained previously
in the metric formulation (\ref{eq:ADS2 solution}).

\subsubsection{Bekenstein-Hawking entropy}

Now let us examine the Bekenstein--Hawking entropy of an extremal
RN black hole in the large $D$ limit. We begin by recalling the $D$-dimensional
RN black hole metric (\ref{eq:metric 1}):

\begin{equation}
ds^{2}=-f\left(r\right)dt^{2}+\frac{dr^{2}}{f\left(r\right)}+r^{2}d\Omega_{n+1}^{2},\label{eq:metric 1-1}
\end{equation}

\noindent where $D=n+3$, and

\begin{equation}
f\left(r\right)=1-\frac{2m}{r^{n}}+\frac{q^{2}}{r^{2n}},\qquad A_{t}=\sqrt{\frac{n+1}{2n}}\left(\frac{q}{r_{+}^{n}}-\frac{q}{r^{n}}\right)dt,
\end{equation}

\noindent where the gauge potential is fixed by requiring $A_{t}\left(r_{+}\right)=0$.
The Bekenstein--Hawking entropy in $D$ dimensions is given by

\begin{equation}
S_{\mathrm{BH}}=\frac{\mathrm{Area}_{D}}{4G_{N}^{\left(D\right)}}=\frac{r_{+}^{D-2}\Omega_{D-2}}{4G_{N}^{\left(D\right)}}.
\end{equation}

\noindent where we used the fact that the horizon area is $\mathrm{Area}_{D}=r_{+}^{D-2}\Omega_{D-2}$.
Using the relation between the two-dimensional and $D$-dimensional
Newton constants at large $D$, as derived in equation (\ref{eq:Newton}):

\begin{equation}
G_{N}^{\left(2\right)}=\underset{n\rightarrow\infty}{\lim}\frac{G_{N}^{\left(D\right)}}{\Omega_{n+1}r_{0}^{n+1}},
\end{equation}

\noindent and applying the extremality condition $r_{+}^{n+1}=r_{0}^{n+1}$
from equation (\ref{eq:infinite}), the Bekenstein-Hawking entropy
becomes

\begin{equation}
S_{\mathrm{BH}}=\frac{1}{4G_{N}^{\left(2\right)}}.\label{eq:large D RN entropy}
\end{equation}

\noindent Therefore, the result follows form ollows from integrating
out the transverse sphere in the dimensional reduction ansatz. The
factor $\Omega_{n+1}r_{0}^{n+1}$ corresponds precisely to the horizon
area contribution of the internal sphere, which becomes renormalized
into the effective gravitational coupling of the two-dimensional theory.

\subsection{CQM$_{1}$ entanglement entropy}

Recall the entanglement entropy of the TFD state in the CFT$_{2}$
obtained previously (\ref{eq:TFD EE}):

\begin{equation}
S_{\mathrm{vN}}\left(A:B\right)=\frac{2\pi r_{+}}{4G_{N}^{\left(3\right)}}.
\end{equation}

\noindent It is straightforward to obtain the entanglement entropy
of the CQM$_{1}$ using the well-known dimensional reduction formula
for Newton\textquoteright s constant:

\begin{equation}
\frac{1}{G_{N}^{\left(2\right)}}=\frac{2\pi r_{+}}{G_{N}^{\left(3\right)}}.
\end{equation}

\noindent Therefore, we find

\begin{equation}
S_{\mathrm{vN}}\left(A:B\right)=\frac{1}{4G_{N}^{\left(2\right)}},\label{eq:CQM1 EE}
\end{equation}

\noindent which exactly reproduces the Bekenstein--Hawking entropy
(\ref{eq:large D RN entropy}) of the large $D$ extremal RN black
hole. It is important to emphasize that extremality is crucial in
our analysis. The emergence of an AdS$_{2}$ factor in the near-horizon
region is a universal property of extremal black holes. This structure
ensures the decoupling of the near-horizon throat from the asymptotic
region and leads to a two-boundary Penrose diagram, thereby making
the AdS$_{2}$/CQM$_{1}$ framework applicable. Without extremality,
the near-horizon geometry does not factorize into $\mathrm{AdS_{2}\times S^{D-2}}$.
Instead, it is approximately $\mathrm{Rindler\times S^{D-2}}$, and
the entropy contains non-universal thermal contributions associated
with finite temperature excitations. In such cases, one does not expect
the Bekenstein--Hawking entropy to be entirely captured by pure inter-boundary
entanglement in a CQM$_{1}$ description. Furthermore, in time-dependent
or dynamical situations, the appropriate framework would be the covariant
Hubeny--Rangamani--Takayanagi (HRT) prescription rather than the
static RT surface. Whether the present identification between black
hole entropy and inter-boundary entanglement entropy can be extended
to the covariant case remains an interesting direction for future
investigation.

This result can also be verified holographically, following the method
of ref.\, \cite{Azeyanagi:2007bj}. Let us recall the two-dimensional
gravity action in Euclidean signature, which arises from the large
$D$ limit of Einstein--Maxwell theory: 

\begin{equation}
I_{\mathrm{string}}=-\frac{1}{16\pi G_{2}}\int d^{2}x\sqrt{\mathbb{G}}e^{-2\phi}\left(\mathbb{R}+4\lambda^{2}-F^{2}\right),\label{eq:JT}
\end{equation}

\noindent where $\lambda=\frac{n}{2r_{0}}$. Using the classical solutions
$\phi=\phi_{0}=\mathrm{const}.$, $A_{t}=\sqrt{\frac{1}{2}}\frac{n}{r_{0}}\left(\rho-\frac{1}{2}\right)d\tau,$
and the field strength $F^{2}=-\frac{4\lambda^{2}q^{2}}{\bar{\mathrm{R}}^{2}}=-4\lambda^{2}$,
one introduces the replica geometry via the curvature singularity
$\mathbb{R}=4\pi\left(1-N\right)\delta^{2}\left(x\right)$. This leads
to the action: $I_{\mathrm{string}}=\frac{N-1}{4G_{N}^{\left(2\right)}}$.
The entanglement entropy is then computed by

\begin{equation}
S_{\mathrm{EE}}=-\frac{\partial}{\partial N}\log\left.\left(e^{-I_{\mathrm{string}}+NI_{\mathrm{string}}^{\left(0\right)}}\right)\right|_{n=1}=\frac{1}{4G_{N}^{\left(2\right)}},
\end{equation}

\noindent where $I_{\mathrm{string}}^{\left(0\right)}$ denotes the
action of the single-sheeted geometry without any branch cut. This
holographic result confirms the field theory calculation (\ref{eq:CQM1 EE}).

In this work, the large $D$ reduction of Einstein--Maxwell theory
leads to the two-dimensional heterotic string effective action (\ref{eq:JT}).
In the extremal limit, the near-horizon solution reduces to a two-dimensional
dilaton gravity theory equivalent to Jackiw--Teitelboim (JT) gravity
without the topological term \cite{Sybesma:2022nby}. In standard
JT gravity, the black hole entropy is determined by the value of the
dilaton at the horizon. In our setup, after taking the extremal and
large $D$ limits, the dilaton profile becomes constant $\phi_{0}$
in the near horizon region, so that $\exp\left(-2\phi_{0}\right)$
acts as an overall multiplicative factor in the action. This constant
factor effectively renormalizes the two-dimensional Newton constant,
yielding the effective coupling $G_{N}^{\left(2\right)}$. Consequently,
the entropy reduces to $S=1/4G_{N}^{\left(2\right)}$, which matches
the standard JT gravity expression when the horizon dilaton value
is identified with the constant background. This result is consistent
with the nearly-AdS$_{2}$/SYK correspondence \cite{Chen:2019qqe},
where the entropy is dominated by the infrared AdS$_{2}$ region and
exhibits universal behavior governed by effective two-dimensional
dilaton gravity. In particular, the large $D$ reduction isolates
precisely this IR AdS$_{2}$ sector, explaining why the entropy is
fully captured by the two-dimensional theory. We emphasize that our
analysis relies on this effective two-dimensional gravitational description
(\ref{eq:JT}) obtained from the large $D$ reduction. The consistency
of such AdS$_{2}$ dilaton gravity frameworks has been extensively
validated in JT gravity and related models, providing a solid conceptual
foundation for applying the entanglement-based interpretation in the
present context.

Finally, we would like to place our results within the broader program
of entanglement-based interpretations of black hole entropy. In our
recent work \cite{Wu:2025qwc}, we identified a deeper mechanism underlying
the statement that entanglement accounts for horizon degrees of freedom.
Specifically, the entanglement entropy admits an effective two-dimensional
description whose near-coincidence limit reduces to a string worldsheet
theory propagating in a curved background. Requiring quantum consistency
of this worldsheet theory leads to background field equations equivalent
to Einstein gravity in AdS, and necessitates the presence of the spacetime
metric, an antisymmetric Kalb--Ramond two-form field, and a dilaton.

In the standard holographic framework, entanglement entropy is computed
geometrically via the RT surface and is therefore expressed in terms
of the spacetime metric. However, within the equivalent string-theoretic
description, the same quantity can be reformulated in terms of a divergenceless
Kalb--Ramond two-form flux. In our construction, the associated conserved
charge flow---sourced by ensembles of open strings---reproduces
known bit-thread configurations and saturates the max-flow/min-cut
bound. This establishes an explicit equivalence between entanglement
entropy and open string charge flux.

Through open--closed string duality, this open string description
is dual to a closed string sector that governs the bulk gravitational
dynamics. In this sense, the Bekenstein--Hawking entropy, geometrically
given by the area law, admits an equivalent interpretation in terms
of closed string charge. The present work provides a concrete realization
of this open closed correspondence in a holographic setting where
the entropy can be computed in a fully controlled manner.

This perspective is conceptually aligned with earlier ideas of Susskind
and Uglum \cite{Susskind:1994sm}, who proposed that black hole entropy
arises from quantum degrees of freedom near the horizon. Here, that
intuition is embedded within a precise holographic framework in which
horizon entropy is explicitly identified with entanglement encoded
in string-theoretic degrees of freedom.

\section{Conclusion and discussion}

In this paper, we developed a method to probe the Bekenstein--Hawking
entropy of black holes via entanglement entropy. This approach is
based on two key observations. On the gravitational side, the near-horizon
geometry of extremal black holes is AdS$_{2}$, and the Bekenstein--Hawking
entropy is entirely determined by this two-dimensional geometry. The
higher-dimensional spherical part of the black hole metric is absorbed
into the $D$-dimensional Newton\textquoteright s constant $G_{N}^{\left(D\right)}$,
which can be effectively reduced to a two-dimensional Newton\textquoteright s
constant $G_{N}^{\left(2\right)}$. On the field theory side, the
entanglement entropy of two disconnected CQM$_{1}$ systems corresponds
to the same AdS$_{2}$ geometry. According to the RT prescription,
this entanglement entropy computes the area of a minimal surface.
Since the near-horizon region of the black hole and the emergent spacetime
derived from entanglement share the same Penrose diagram---with both
the black hole event horizon and the RT surface corresponding to specific
points on this diagram---the Bekenstein--Hawking entropy can be
extracted from entanglement entropy when these points coincide.

We explicitly verified this correspondence in three examples: the
BTZ black hole, the $D1$-$D5$ black hole in type IIB string theory,
and the large $D$ RN black hole.

Our results suggest the following:
\begin{itemize}
\item Entanglement across the event horizon is the origin of the Bekenstein--Hawking
entropy. In other words, information in $D$ spacetime dimensions
can be encoded within a one-dimensional quantum system.
\item Entanglement entropy provides a route to microscopically count the
BPS states of black holes. This is because the RT formula bridges
quantum features (states) and classical geometry (minimal surfaces),
and the minimal surface in the near-horizon region corresponds to
the black hole event horizon.
\end{itemize}
For future work, a promising direction is:
\begin{itemize}
\item Extension to the covariant case. It would be important to generalize
this method to time-dependent or non-static geometries, potentially
involving the covariant holographic entanglement entropy framework.
\end{itemize}
\noindent \bigskip 

\vspace{5mm}

\noindent {\bf Acknowledgements} 
This work was supported by NSFC Grant No. 12105031 and No. 12347101.


\begin{thebibliography}{99}

\bibitem{Susskind:1994sm} L.~Susskind and J.~Uglum, ``Black hole entropy in canonical quantum gravity and superstring theory,'' Phys. Rev. D \textbf{50}, 2700-2711 (1994) doi:10.1103/PhysRevD.50.2700 [arXiv:hep-th/9401070 [hep-th]]. 



\bibitem{Fiola:1994ir} T.~M.~Fiola, J.~Preskill, A.~Strominger and S.~P.~Trivedi, ``Black hole thermodynamics and information loss in two-dimensions,'' Phys. Rev. D \textbf{50}, 3987-4014 (1994) doi:10.1103/PhysRevD.50.3987 [arXiv:hep-th/9403137 [hep-th]]. 


\bibitem{Jacobson:1994iw} T.~Jacobson, ``Black hole entropy and induced gravity,'' [arXiv:gr-qc/9404039 [gr-qc]]. 

\bibitem{Azeyanagi:2007bj} T.~Azeyanagi, T.~Nishioka and T.~Takayanagi, ``Near Extremal Black Hole Entropy as Entanglement Entropy via AdS(2)/CFT(1),'' Phys. Rev. D \textbf{77}, 064005 (2008) doi:10.1103/PhysRevD.77.064005 [arXiv:0710.2956 [hep-th]]. 




\bibitem{Emparan:2006ni} R.~Emparan, ``Black hole entropy as entanglement entropy: A Holographic derivation,'' JHEP \textbf{06}, 012 (2006) doi:10.1088/1126-6708/2006/06/012 [arXiv:hep-th/0603081 [hep-th]]. 



\bibitem{Emparan:1999wa} R.~Emparan, G.~T.~Horowitz and R.~C.~Myers, ``Exact description of black holes on branes,'' JHEP \textbf{01}, 007 (2000) doi:10.1088/1126-6708/2000/01/007 [arXiv:hep-th/9911043 [hep-th]]. 










\bibitem{Maldacena:2001kr} J.~M.~Maldacena, ``Eternal black holes in anti-de Sitter,'' JHEP \textbf{04}, 021 (2003) doi:10.1088/1126-6708/2003/04/021 [arXiv:hep-th/0106112 [hep-th]]. 





\bibitem{Cardy:2016fqc} J.~Cardy and E.~Tonni, ``Entanglement hamiltonians in two-dimensional conformal field theory,'' J. Stat. Mech. \textbf{1612}, no.12, 123103 (2016) doi:10.1088/1742-5468/2016/12/123103 [arXiv:1608.01283 [cond-mat.stat-mech]]. 

\bibitem{Jiang:2024ijx} X.~Jiang, P.~Wang, H.~Wu and H.~Yang, ``Alternative to purification in conformal field theory,'' Phys. Rev. D \textbf{111}, no.2, L021902 (2025) doi:10.1103/PhysRevD.111.L021902 [arXiv:2406.09033 [hep-th]]. 


\bibitem{Jiang:2024xqz} X.~Jiang, P.~Wang, H.~Wu and H.~Yang, ``Realization of ''ER=EPR'','' [arXiv:2411.18485 [hep-th]]. 




\bibitem{Sen:2008vm} A.~Sen, ``Quantum Entropy Function from AdS(2)/CFT(1) Correspondence,'' Int. J. Mod. Phys. A \textbf{24}, 4225-4244 (2009) doi:10.1142/S0217751X09045893 [arXiv:0809.3304 [hep-th]]. 





\bibitem{Emparan:2013xia} R.~Emparan, D.~Grumiller and K.~Tanabe, ``Large-D gravity and low-D strings,'' Phys. Rev. Lett. \textbf{110}, no.25, 251102 (2013) doi:10.1103/PhysRevLett.110.251102 [arXiv:1303.1995 [hep-th]]. 



\bibitem{Guo:2015swu} E.~D.~Guo, M.~Li and J.~R.~Sun, ``CFT dual of charged AdS black hole in the large dimension limit,'' Int. J. Mod. Phys. D \textbf{25}, no.07, 1650085 (2016) doi:10.1142/S0218271816500851 [arXiv:1512.08349 [gr-qc]]. 

\bibitem{Tong:2023nvi} C.~W.~Tong, D.~H.~Du and J.~R.~Sun, ``Island of Reissner-Nordstr\"om anti\textendash{}de Sitter black holes in the large D limit,'' Phys. Rev. D \textbf{109}, no.10, 104053 (2024) doi:10.1103/PhysRevD.109.104053 [arXiv:2306.06682 [hep-th]]. 

\bibitem{Sybesma:2022nby} W.~Sybesma, ``A zoo of deformed Jackiw-Teitelboim models near large dimensional black holes,'' JHEP \textbf{01}, 141 (2023) doi:10.1007/JHEP01(2023)141 [arXiv:2211.07927 [hep-th]]. 





\bibitem{Hartman:2015}
T.~Hartman, ``Lectures on Quantum Gravity and Black Holes.''

\bibitem{Callebaut:2023fnf} N.~Callebaut, ``Entanglement in Conformal Field Theory and Holography,'' Lect. Notes Phys. \textbf{1022}, 239-271 (2023) doi:10.1007/978-3-031-42096-2\_10 [arXiv:2303.16827 [hep-th]]. 






\bibitem{Maldacena:1998bw} J.~M.~Maldacena and A.~Strominger, ``AdS(3) black holes and a stringy exclusion principle,'' JHEP \textbf{12}, 005 (1998) doi:10.1088/1126-6708/1998/12/005 [arXiv:hep-th/9804085 [hep-th]]. 

\bibitem{Strominger:1996sh} A.~Strominger and C.~Vafa, ``Microscopic origin of the Bekenstein-Hawking entropy,'' Phys. Lett. B \textbf{379}, 99-104 (1996) doi:10.1016/0370-2693(96)00345-0 [arXiv:hep-th/9601029 [hep-th]]. 


\bibitem{Callan:1996dv} C.~G.~Callan and J.~M.~Maldacena, ``D-brane approach to black hole quantum mechanics,'' Nucl. Phys. B \textbf{472}, 591-610 (1996) doi:10.1016/0550-3213(96)00225-8 [arXiv:hep-th/9602043 [hep-th]]. 













\bibitem{Myers:1986un} R.~C.~Myers and M.~J.~Perry, ``Black Holes in Higher Dimensional Space-Times,'' Annals Phys. \textbf{172}, 304 (1986) doi:10.1016/0003-4916(86)90186-7 


\bibitem{McGuigan:1991qp} M.~D.~McGuigan, C.~R.~Nappi and S.~A.~Yost, ``Charged black holes in two-dimensional string theory,'' Nucl. Phys. B \textbf{375}, 421-450 (1992) doi:10.1016/0550-3213(92)90039-E [arXiv:hep-th/9111038 [hep-th]]. 


\bibitem{Giveon:2005jv} 
A.~Giveon and D.~Kutasov, ``The Charged black hole/string transition,'' JHEP \textbf{01}, 120 (2006) doi:10.1088/1126-6708/2006/01/120 [arXiv:hep-th/0510211 [hep-th]]. 


 
\bibitem{Chen:2019qqe} Y.~Chen and P.~Zhang, ``Entanglement Entropy of Two Coupled SYK Models and Eternal Traversable Wormhole,'' JHEP \textbf{07}, 033 (2019) doi:10.1007/JHEP07(2019)033 [arXiv:1903.10532 [hep-th]]. 


\bibitem{Wu:2025qwc} H.~Wu and S.~Ying, ``Towards a worldsheet theory of entanglement entropy,''  accepted by Phys. Rev. D, doi:10.1103/cgg9-p6xy [arXiv:2511.16586 [hep-th]].  









\end{thebibliography}
\end{document}